\documentclass[%
 reprint,
 amsmath,amssymb,
 aps,
nofootinbib]{revtex4-1}
\usepackage{scrextend}
\usepackage{graphicx}
\usepackage{dcolumn}
\usepackage{bm}
\usepackage{float}
\usepackage{hyperref}
\usepackage[mathlines]{lineno}
\usepackage{footnote}
\usepackage{amsmath}
\usepackage{xfrac}
\usepackage{braket}
\usepackage{cancel}
\usepackage{xcolor}
\usepackage{amssymb}

\usepackage{mathtools}

\newcommand{\arXiv}[2]{\href{http://arxiv.org/pdf/hep-ph/#1}{{\tt #2/#1}}}
\newcommand{\arXivold}[2]{\href{http://arxiv.org/pdf/#1}{{\tt #2/#1}}}

\usepackage{blindtext} 
\newcommand{\subf}[2]{%
  {\small\begin{tabular}[t]{@{}c@{}}
  #1\\#2
  \end{tabular}}%
}

\begin{document}

\preprint{APS/123-QED}

\title{Model-independent Veltman Condition, Naturalness and the Little Hierarchy Problem}
\author{Fayez Abu-Ajamieh}
 \email{fayez.abu-ajamieh@umontpellier.fr}
\affiliation{%
LUPM UMR5299, Universit\'e de Montpellier, 34095 Montpellier, France
}


\begin{abstract}
We adopt a bottom-up Effective Field Theory (EFT) approach to derive a model-independent Veltman condition to cancel out the quadratic divergences in the Higgs mass. We show using the equivalence theorem that all the deviations in the Higgs couplings to the $W$ and $Z$ from the SM predictions should vanish. We argue based on tree-level unitarity that any new physics that naturally cancels out the quadratic divergences should be $\lesssim 19$ TeV. We show that the level of fine-tuning required is $O(0.1-1\%)$ unless the UV sector has a symmetry that forces the satisfaction of the model-independent Veltman condition, in which case all fine-tuning is eliminated. We also conjecture that, if no new physics that couples to the Higgs is observed up to $\sim 19$ TeV, or if the Higgs couplings to the SM particles conform to the SM predictions, then the Higgs either does not couple to any UV sector or is fine-tuned.
\end{abstract}

\pacs{Valid PACS appear here}
\maketitle


\section{\label{sec:Introduction}Introduction}
The discovery of the Higgs boson at the LHC represents a resounding triumph of the Standard Model (SM) of elementary particles. With this discovery, the final piece of the SM has been put in place, and we now have a UV complete theory that can provide accurate predictions that can be measured with high precision at the LHC. Nonetheless, it is almost certain that the SM is not the full story, and that new physics Beyond the Standard Model (BSM) is required. There is still no accepted theory of quantum gravity, no agreed-upon explanation of neutrino masses, dark matter, or matter-antimatter asymmetry, and no explanation of dark energy. Furthermore, the SM itself suffers from inconsistencies that appear to suggest the need for new physics. The hierarchy problem, which refers to the sensitivity of the Higgs mass to UV corrections, remains one of the most crucial inconsistencies in the SM that are yet to be solved.

\begin{figure}[ht] 
\centering
\includegraphics[width=0.5\textwidth]{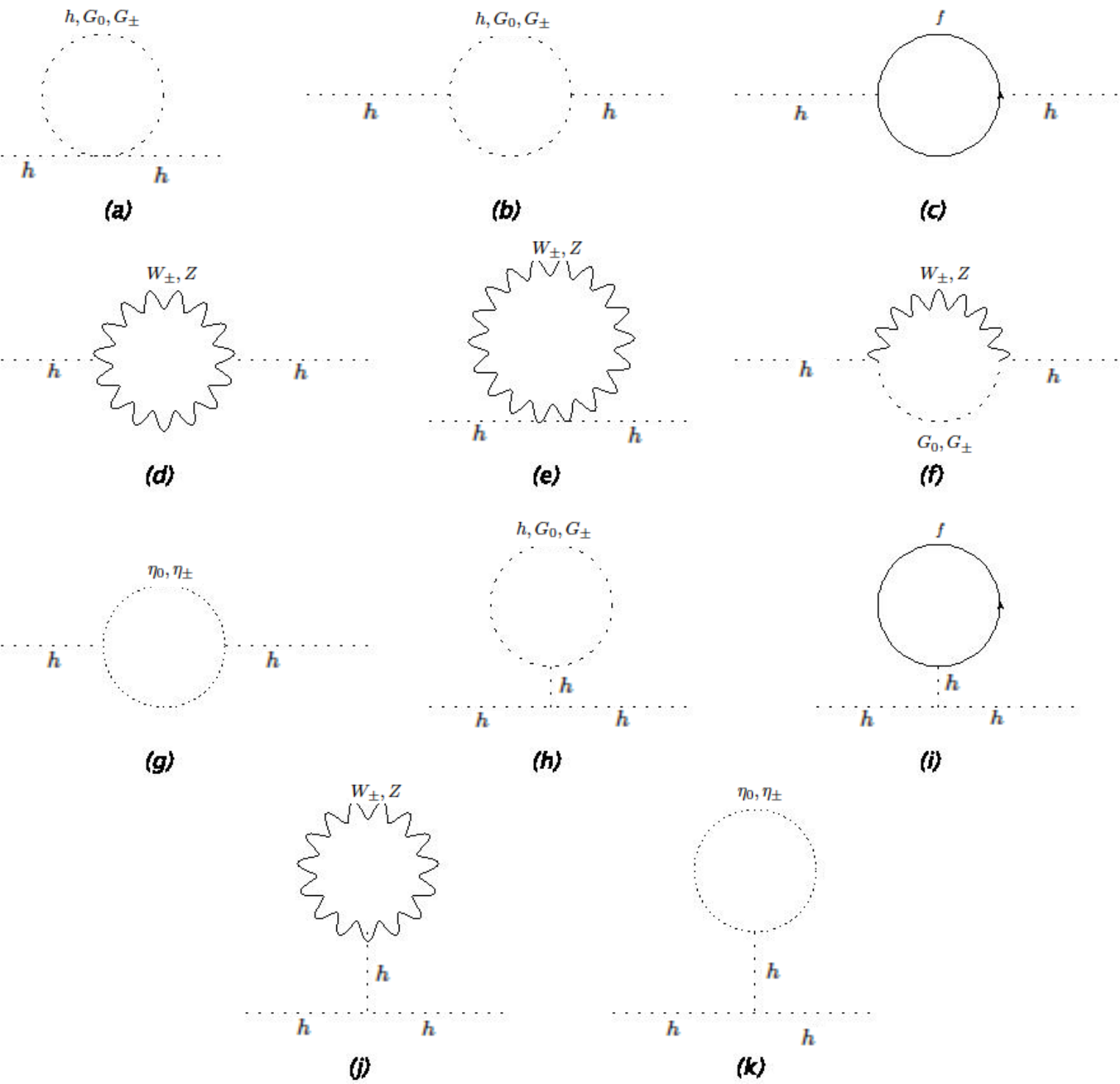}
\caption{1-loop Higgs mass corrections in the Landau Gauge. Here, $G_{0}$ and $G_{\pm}$ are the longitudinal modes of the massive gauge bosons, whereas $W^{\pm}$ and $Z$ are the transverse modes. }
\label{fig:Feynman1}
\end{figure}

It was first pointed out in \cite{Susskind:1978ms} that a theory with a fundamental scalar at the Electroweak (EW) scale would require the fine-tuning of $O(10^{-34})$ in the scalar's self-energy, thereby making the theory unnatural. The concept of naturalness states that, for a theory to be natural, it shouldn't be too sensitive to the fundamental constants of nature at ordinary energies \cite{Susskind:1982mw}. When applied to the SM Higgs, the 1-loop Higgs mass corrections (see Fig. \ref{fig:Feynman1}) are given by \cite{Veltman:1980mj}
\begin{equation}\label{eq:SM_Veltman}
\delta m_{h}^{2} = \frac{3\Lambda^{2}}{8\pi^{2} v^{2}}\Big[ 4m_{t}^{2} - 2m_{W}^{2} - m_{Z}^{2} - m_{h}^{2} \Big] + O(\log{\frac{\Lambda^{2}}{m_{h}^{2}}}),
\end{equation}
where $\Lambda$ is a UV cutoff scale, $v$ is the Higgs VEV, and only the mass of the top quark is retained out of all fermions. For the Higgs mass to remain at the EW scale and thus be natural, one has to tune $\delta m_{h}^{2} \simeq 0$. This is the so-called Veltman condition. 
Prior to the discovery of the top quark and the Higgs boson, the Veltman condition was used to make (wrong) predictions for their masses \cite{CapdequiPeyranere:1990gk, LopezCastro:1994mx}. However, as the masses of the top quark and the Higgs boson became known, it became apparent that the Veltman condition could not be satisfied as it stands. This led to some efforts to explain the hierarchy problem as an artifact of the renormalization scheme \cite{Bardeen:1995kv, Grange:2013yjn, Aoki:2012xs, Vieira:2012ex, Manohar:2018aog}; however, this can only be accurate if there is no new physics that couples to the Higgs sector in the UV, as the masses of such heavy states should appear in the low energy theory.

The modern view of the hierarchy problem is that, when the Higgs sector is treated as an EFT, then its mass suffers from quadratic sensitivity to the scale at which the UV theory comes into play (for instance, see \cite{falkowski} for a pedagogical introduction to the topic). More specifically, if the Higgs boson couples to some heavy sector at a UV scale $\Lambda$, then after integrating out the heavy degrees of freedom, one finds that the Higgs mass is schematically given by
\begin{equation}\label{eq:EFT_higgs}
m_{h}^{2} = m_{H}^{2} - \frac{1}{16\pi^{2}}\Bigg[ C_{0} \Lambda^{2} + C_{1} m_{H}^{2} + C_{2}\frac{m_{H}^{4}}{\Lambda^{2}} + \cdots \Bigg],
\end{equation}
where $m_{h}$ is the physical Higgs mass, $m_{H}$ is the bare mass, and $C_{i}$ represent Wilson coefficients. To have $m_{h}^{2}$ at the EW scale, one has to tune $m_{H}^{2}$ against the second term $\sim O(\Lambda^{2})$. This makes the physical Higgs mass sensitive to whatever UV completion that has been integrated out, rendering the Higgs sector unnatural. 

However, if there is no UV sector that couples to the Higgs, then there will be no heavy degrees of freedom to integrate out in the first place, and  Eq. \ref{eq:EFT_higgs} will be free of any terms $\sim \Lambda^{2}$, thus eliminating the issue of fine-tuning and making the Higgs mass natural. In this case, the hierarchy problem will simply reduce to a mere artifact of the renormalization scheme used and can be resolved by formulating an appropriate scheme, or by simply using dimensional regularization and subtracting the suitable counterterms. Nonetheless, there are strong reasons to believe that the Higgs boson should couple to new physics in the UV, most notable of which is to provide a natural explanation for neutrino masses.\footnote{One can, of course, assume that the neutrino masses are of the Dirac type and thus avoid the need for UV physics such as the seesaw mechanism, however, fine-tuning will simply reappear in the minuscule Yukawa couplings of the active neutrinos, and thus the issue of fine-tuning in the SM remains, albeit in a different form.}

In this study, we tackle the issue of the Higgs mass naturalness and the hierarchy problem using a completely model-independent bottom-up approach. We use an EFT Lagrangian to calculate the 1-loop corrections to the Higgs mass and derive a model-independent Veltman condition in terms of the deviations in the Higgs couplings from the SM predictions. Similar work has been undertaken in \cite{Biswas:2020abl} within the context of the Standard Model EFT (SMEFT), where the dimension-6 SMEFT operators were used to derive the modified Veltman condition. In \cite{Chakraborty:2016izr}, a bottom-up approach was used to argue that new scalar degrees of freedom that couple both to the Higgs boson and to fermions should exist. However, in this study, we follow a completely model-independent approach based on the results presented in \cite{Chang:2019vez,Abu-Ajamieh:2020yqi}, and we remain completely agnostic with regard to any UV completion.

The self-couplings of the Higgs boson and its couplings to the remaining SM particles are currently being measured at the LHC. Although these measurements are consistent with the SM predictions, they nonetheless leave ample room for potential deviations from the SM predictions, and thus, for the possibility of new physics BSM. For instance, the couplings of the Higgs to other SM particles are only measured at $O(20\%)$, whereas the Higgs self-coupling is only constrained to be $\lesssim 10$. Thus, the effective Lagrangian can be written as
\begin{flalign}\label{eq:UV_Lag_Unitary}
\mathcal{L} & = \mathcal{L}_{\text{SM}} -\delta_{3}\frac{m_{h}^{2}}{2v}h^{3}-\delta_{4}\frac{m_{h}^{2}}{8v^{2}}h^{4} - \sum_{n=5}^{\infty}\frac{c_{n}}{n!}\frac{m_{h}^{2}}{v^{n-2}}	h^{n} \nonumber \\
& + \delta_{Z1}\frac{m_{Z}^{2}}{v}hZ_{\mu}Z^{\mu} + \delta_{W1}\frac{2m_{W}^{2}}{v}hW_{\mu}^{+}W^{\mu-} \nonumber \\
& + \delta_{Z2}\frac{m_{Z}^{2}}{2v^{2}}h^{2}Z_{\mu}Z^{\mu} + \delta_{W2}\frac{m_{W}^{2}}{v^{2}}h^{2}W_{\mu}^{+}W^{\mu-} \nonumber \\
& + \sum_{n=3}^{\infty}\Bigg[ \frac{c_{Zn}}{n!}\frac{m_{Z}^{2}}{v^{n}}h^{n}Z_{\mu}Z^{\mu}+\frac{c_{Wn}}{n!}\frac{2m_{W}^{2}}{v^{n}}h^{n}W_{\mu}^{+}W^{\mu-}\Bigg] \nonumber \\
& -\delta_{t1} \frac{m_{t}}{v}h\bar{t}t -\sum_{n=2}^{\infty}\frac{c_{tn}}{n!}\frac{m_{t}}{v^{n}}h^{n}\bar{t}t + \cdots,
\end{flalign}
where $\delta_{i}$ parametrize the deviations of the Higgs couplings from the SM prediction:
\begin{equation}\label{eq:deviations}
\delta_{i} = \frac{g_{i}- g_{i}^{\text{SM}}}{g_{i}^{\text{SM}}},
\end{equation}
while $c_{i}$ represent Wilson coefficients that do not have SM counterparts. In this parameterization, we expand the BSM operators that could couple to the Higgs in terms of the dimension of the operator without assuming any expansion scale $M$. Here, we scale the Wilson coefficients by the powers of the Higgs VEV to keep them dimensionless. This parameterization is more model-independent than SMEFT. This is because in SMEFT, one assumes a single expansion scale $M$ for the higher-order operators, which need not be the case; in contrast, here, we only assume deviations in the Higgs couplings and only demand that they be consistent with measurements. Furthermore, this parameterization is more transparent, as in the LHC, couplings are measured, not the expansion scale. The ellipses denote operators with higher powers and/or derivatives that do not contribute to the Higgs mass. Note that the BSM couplings and Wilson coefficients are perturbative, as they will be implicitly suppressed by the scale of new physics; see \cite{Chang:2019vez,Abu-Ajamieh:2020yqi} for more detail.

Using the BSM Lagrangian, we can calculate the 1-loop corrections to the Higgs mass and use them to extract the BSM Veltman condition in terms of the deviations in the Higgs couplings. As we show in this paper, given the current accuracy of the Higgs couplings measurements, it is possible to cancel out the UV divergences in the Higgs mass with a scale of new physics as high as $\sim 19$ TeV, thus providing a potential solution to the little hierarchy problem. We show that this can be achieved at the price of $O(0.1-1\%)$ fine-tuning, unless the UV completion has a symmetry that forces the satisfaction of the BSM Veltman condition, in which case the fine-tuning is averted. We also show that there should be no deviations from the SM predictions in the massive gauge boson sector.

The remainder of this paper is organized as follows: In Sec. \ref{Sec:BSM_Veltman}, we review the model-independent approach presented in \cite{Chang:2019vez, Abu-Ajamieh:2020yqi}, and we derive the model-independent BSM Veltman condition. In Sec. \ref{Sec:Maximum_Energy}, we discuss the maximum energy scale of new physics that can be achieved while satisfying the BSM Veltman condition based on tree-level unitarity. In Sec. \ref{Sec:fine-tuning}, we estimate the fine-tuning associated with the BSM Veltman condition and discuss the aspects of naturalness. Finally, we present our conclusions in Sec. \ref{Sec:conclusions}.

\section{Model-independent Veltman Condition}\label{Sec:BSM_Veltman}
\subsection{Review of the Model-independent Approach}\label{Sec:review}
We begin by reviewing the approach presented in \cite{Chang:2019vez, Abu-Ajamieh:2020yqi}. The Lagrangian in Eq. (\ref{eq:UV_Lag_Unitary}) represents the most general BSM interactions with the Higgs boson in the unitary gauge.\footnote{In principle, it is possible to extend Eq. (\ref{eq:UV_Lag_Unitary}) by including operators with more derivatives, however, as argued in \cite{Abu-Ajamieh:2020yqi}, this would only enhance the energy growth of the amplitudes and thus lower the scale of new physics. Here we choose to be conservative and neglect operators with more derivatives.} In our calculation, we would like to be able to use the equivalence theorem; thus it is more convenient to work in a gauge where the Goldstone bosons are manifest. Thus we write the Higgs doublet as
\begin{equation}\label{eq:Higgsdoublet}
H = \frac{1}{\sqrt{2}} 
\begin{pmatrix}
G_{1} + i G_{2}\\
v + h + i G_{0}
\end{pmatrix},
\end{equation}
and then, we define the field
\begin{equation}\label{eq:defineX}
X \equiv \sqrt{2H^{\dagger}H} -v = h + \frac{\vec{G}^{2}}{2(v+h)}-\frac{\vec{G}^{4}}{8(v+h)^{3}}+ \cdots,
\end{equation}
where $\vec{G}^{2} = G_{0}^{2} + G_{1}^{2} +G_{2}^{2}$. As $X = h$ in the unitary gauge, Eq. (\ref{eq:UV_Lag_Unitary}) can be generalized to a general gauge by making the replacement $h \rightarrow X$. Note that $X$ is non-analytic at $H = 0$; however, we are only interested in the region where $\braket{H} \neq 0$. With this replacement, the relevant part of the Higgs potential that contributes to the Higgs mass at $1$-loop takes the form
\begin{equation}\label{eq:BSMHiggsLagrangian}
\delta\mathcal{L}_{h} \supset -\frac{3m_{h}^{2} \delta_{3}}{3!v}h^{3} - \frac{3m_{h}^{2} \delta_{4}}{4!v^{2}}h^{4}-\frac{3m_{h}^{2}\delta_{3}}{4v^{2}}h^{2} \big[2G_{+}G_{-}+ G_{0}^{2}\big],
\end{equation}
where $\delta_{3,4}$ represent the deviations in the Higgs cubic and quartic couplings compared with the SM predictions as defined in Eq. (\ref{eq:deviations}), and $G_{0}$, $G_{\pm} = \frac{1}{\sqrt{2}}(G_{1} \mp i G_{2})$ are the Goldstone bosons corresponding to the longitudinal modes of the $Z$ and $W^{\pm}$, respectively.

To restore the Goldstone bosons in the massive gauge boson sector, we define the gauge-invariant operator 
\begin{equation}\label{eq:WZprojector}
\hat{H} \equiv \frac{H}{\sqrt{H^{\dagger}H}} = \begin{pmatrix}
0\\
1
\end{pmatrix}
+ O(\vec{G}),
\end{equation}
and we make the following replacements in Eq. (\ref{eq:UV_Lag_Unitary})
\begin{align}
\begin{split}\label{eq:WZreplacements1}
 Z_{\mu} \rightarrow \hat{H}^{\dagger} i D_{\mu} \hat{H} = & -\frac{m_{Z}}{v}Z_{\mu} - \frac{1}{v}\partial_{\mu} G_{0} + \cdots 
\end{split}\\
\begin{split}\label{eq:WZreplacements2}
  W_{\mu}^{+} \rightarrow \tilde{\hat{H}}^{\dagger} i D_{\mu} \hat{H} = & \frac{\sqrt{2}m_{W}}{v}W_{\mu}^{+} + \frac{i\sqrt{2}}{v} \partial_{\mu}G_{+} + \cdots 
\end{split}\\
\begin{split}
W_{\mu}^{-} \rightarrow \hat{H}^{\dagger}i D_{\mu} \tilde{\hat{H}} = & \frac{\sqrt{2}m_{W}}{v}W_{\mu}^{-} - \frac{i\sqrt{2}}{v} \partial_{\mu}G_{-} +\cdots\label{eq:WZreplacements3}
\end{split}
\end{align}
where $\tilde{\hat{H}} = \epsilon \hat{H}^{*}$ and $\epsilon$ is the anti-symmetric $2 \times 2$ tensor. With these replacements, the relevant part of the BSM massive gauge boson sector is expressed as	
\begin{flalign}\label{eq:BSMMassiveLagrangian}
\delta\mathcal{L}_{W/Z} & \supset \frac{\delta_{Z1}}{v}h (\partial G_{0})^{2} + \frac{(\delta_{Z2} - 4\delta_{Z1})}{2v^{2}}h^{2} (\partial G_{0})^{2} & \nonumber \\ 
& -\frac{2\delta_{Z1}}{v^{2}}G_{0} h \partial_{\mu}G_{0}\partial^{\mu}h +\frac{2\delta_{W1}}{v}h\partial_{\mu}G_{+}\partial^{\mu}G_{-} & \nonumber\\
& + \frac{(\delta_{W2}-4\delta_{W1})}{v^{2}} h^{2} \partial_{\mu}G_{+}\partial^{\mu}G_{-} + \frac{m_{Z}^{2}\delta_{Z1}}{v}hZ_{\mu}Z^{\mu} & \nonumber \\
& + \frac{m_{W}^{2} \delta_{W2}}{v^{2}}h^{2}W_{\mu}^{+}W^{-\mu}  + \frac{2m_{Z}\delta_{Z1}}{v}hZ_{\mu}\partial^{\mu}G_{0} & \nonumber\\ 
& + \frac{m_{Z}^{2} \delta_{Z2}}{2v^{2}}h^{2}Z_{\mu}Z^{\mu} -\frac{2\delta_{W1}}{v^{2}}h \partial^{\mu}h\Big[G_{-}\partial_{\mu}G_{+} & \nonumber \\
 & +G_{+} \partial_{\mu}G_{-} \Big] + \frac{2m_{W}\delta_{W1}}{v}h \Big[ m_{W}W_{\mu}^{+}W^{-\mu} & \nonumber \\
& + i \Big(W_{\mu}^{-} \partial^{\mu}G_{+}-W_{\mu}^{+} \partial^{\mu}G_{-} \Big) \Big] .
\end{flalign}

Note that we are not assuming custodial symmetry; hence, $\delta_{W1(2)}$ could, in principle, be different from $\delta_{Z1(2)}$. However, as the experimental limits on the $\rho$ parameter are extremely tight \cite{Tanabashi:2018oca}
\begin{equation}\label{eq:CustodialCorrection}
|\rho| = 1.00039 \pm 0.00019,
\end{equation}
it is only meaningful to consider the cases where $\rho = \pm 1$ \cite{Stolarski:2020qim}\footnote{Notice here that while the SM prediction is $\rho =+1$, collider searches are only sensitive to the magnitude of $\rho$ and not its sign. Thus, it is possible in principle to have negative values of $\rho$. We refer the interested reader to the indicated reference for a detailed study of this issue.}, in which case $\delta Z_{1(2)} = \pm \delta W_{1(2)}$. We refer to $\rho=1$ as the custodial limit, while $\rho = -1$ is referred to as the anti-custodial limit.

To restore the Goldstone dependence in the fermion sector, we write the Higgs interaction with fermions as
\begin{equation}\label{eq:BSMfermion}
\delta \mathcal{L}_{f} = -m_{f}( \bar{Q}_{L}\tilde{\hat{H}}q_{R} + \text{h.c.})\Big(\delta_{f1}\frac{X}{v} + c_{f2} \frac{X^{2}}{2!v^{2}} +\cdots \Big).
\end{equation}

In the remainder of this paper, we only retain the top quark, as it has the largest contribution to the Higgs mass. However, including the remaining SM fermions is straightforward. The relevant part of the top sector thus is expressed as
\begin{equation}\label{eq:BSMtop}
\delta \mathcal{L}_{\text{top}} \supset - \frac{m_{t} \delta_{t1}}{v}h\bar{t}t -\frac{m_{t}c_{t2}}{2v^{2}}h^{2}\bar{t}t.
\end{equation}

Finally, in principle, one could include the corrections to the Higgs mass arising from its couplings to massless gauge bosons; however, as the Higgs only couples to photons and gluons through loops, corrections to the Higgs mass only begin at 2 and 3 loops, and thus, can be neglected. Throughout this paper, we limit our calculation to 1-loop.

\subsection{BSM Loop Corrections and the Model-independent Veltman Condition}\label{sec:BSMcondition}
\begin{figure}[H] 
  \centering
    \includegraphics[width=0.17\textwidth]{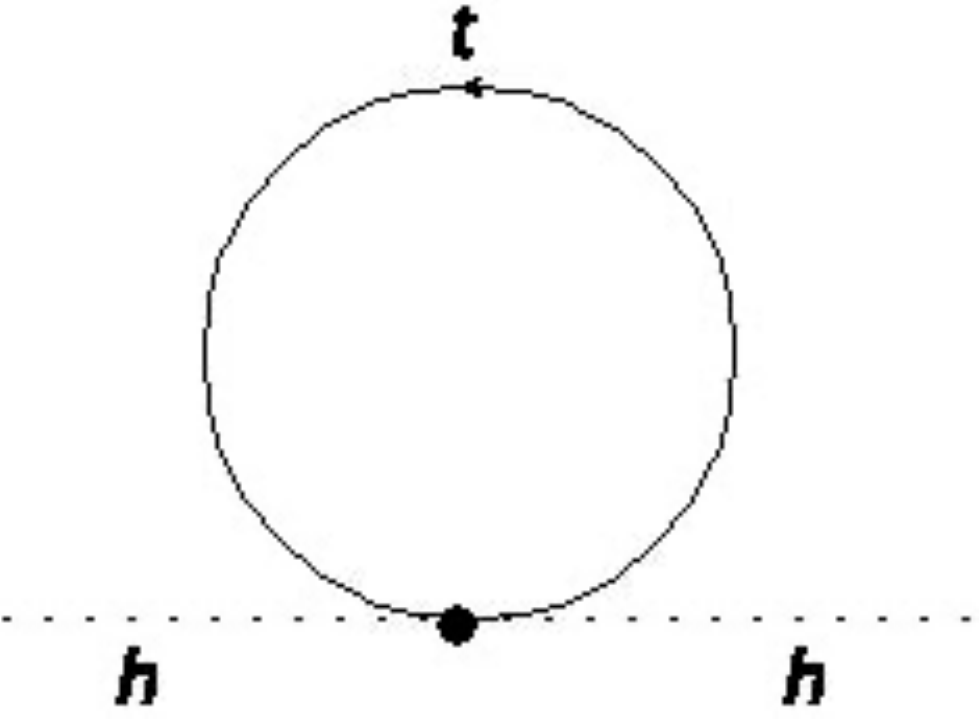}
      \caption{Additional top loop contribution to the mass of the Higgs. This originates from the second term in Eq. \ref{eq:BSMtop}.}
      \label{fig:Feynman2}
\end{figure}

We can now calculate the BSM loop corrections to the Higgs mass. Note that the BSM contributions given in Eqs. (\ref{eq:BSMHiggsLagrangian}), (\ref{eq:BSMMassiveLagrangian}), and (\ref{eq:BSMtop}) are to be added to the SM contributions. The corrections to the Higgs mass are shown in Fig. \ref{fig:Feynman1}. There is another contribution from the effective coupling $c_{t2}$ (the second term in Eq. \ref{eq:BSMtop}) shown in Fig. \ref{fig:Feynman2}. Note also that, if there are heavy degrees of freedom that couple to the Higgs, then they too will contribute to the Higgs mass at 1-loop; however, in the low energy theory, the part of these contributions that is not encoded in the deviations or Wilson coefficients should be suppressed by $\Lambda^{2}$, and thus can be neglected.
\newline
In our calculation, we work in the Landau gauge $\xi = 0$ and use a UV-cutoff regularization scheme, as it is more suitable for predicting the scale of new physics. We avoid using dimensional regularization because, when using it, the scale of any potential new physics is obscured. We discuss this point in more detail in the following section. In the Landau gauge, all ghost contributions (diagrams (g) and (k) in Fig \ref{fig:Feynman1}) vanish. 

Before we proceed with finding the BSM Veltman condition, let us first focus on the contributions from the longitudinal modes, i.e., the Feynman diagrams in (a), (b), (f), and (h) in Fig. \ref{fig:Feynman1}. The diagram in (f) provides a vanishing contribution in the Landau gauge, whereas the other ones provide contributions that are quartically divergent:
\begin{flalign}
\mathcal{M}_{G_{0}}^{(a)} & =  -\frac{1}{32\pi^{2}v^{2}}(\delta_{Z2} - 4\delta_{Z1}) \Lambda^{4} +O(\Lambda^{2}), \tag{\theequation{}a} \label{eq:QuarticDiv-a}\\ 
\mathcal{M}_{G_{\pm}}^{(a)} & =  -\frac{1}{16\pi^{2}v^{2}}(\delta_{W2} - 4\delta_{W1}) \Lambda^{4} +O(\Lambda^{2}), \tag{\theequation{}b} \label{eq:QuarticDiv-b}\\
\mathcal{M}_{G_{0}}^{(b)} & =  -\frac{\delta_{Z1}^{2}}{16\pi^{2}v^{2}}\Lambda^{4} +O(\Lambda^{2}), \tag{\theequation{}c} \label{eq:QuarticDiv-c}\\
\mathcal{M}_{G_{\pm}}^{(b)} & =  -\frac{\delta_{W1}^{2}}{8\pi^{2}v^{2}}\Lambda^{4} +O(\Lambda^{2}),\tag{\theequation{}d} \label{eq:QuarticDiv-d}\\
\mathcal{M}_{G_{0}}^{(h)} & =  \frac{3(1+\delta_{3})\delta_{Z1}}{32\pi^{2}v^{2}}\Lambda^{4} +O(\Lambda^{2}), \tag{\theequation{}e} \label{eq:QuarticDiv-e}\\
\mathcal{M}_{G_{\pm}}^{(h)} & =  \frac{3(1+\delta_{3})\delta_{W1}}{16\pi^{2}v^{2}}\Lambda^{4} +O(\Lambda^{2}). \tag{\theequation{}f} \label{eq:QuarticDiv-f}
\end{flalign}
These quartic divergences arise from the contributions of the longitudinal modes of the massive gauge bosons as a direct result of the equivalence theorem. As each longitudinal mode is replaced with a derivative, each insertion of a longitudinal mode will increase the energy dependence of the mass correction by one power. In the SM, these quartic divergences are guaranteed to cancel out by gauge invariance, which relates the 3-gauge boson and the 4-gauge boson couplings; however, in the EFT, they do not cancel out, as gauge invariance is no longer manifest. More specifically, if we upset the SM predictions by allowing deviations in the couplings of the gauge bosons, then gauge invariance is no longer manifest in the low energy theory (although it should be restored in the full theory), and quartic divergences no longer cancel out. Thus, it is necessary to ensure the vanishing of these quartic divergences first, as they are the leading corrections to the Higgs mass at 1-loop.\footnote{Notice that in an EFT approach, divergences of any power could arise, however, divergences of higher powers only arise beyond the 1-loop order, and at 1-loop, the only quartic divergences arise from Eqs. (\ref{eq:QuarticDiv-a})-(\ref{eq:QuarticDiv-f}).}

Confining ourselves to the custodial and anti-custodial limits, the conditions for canceling out the quartic divergence are expressed as	
\begin{eqnarray}
\delta_{V1}(3\delta_{3} - 2\delta_{V1} + 7) - \delta_{V2} =0,\hspace{12.5mm} \text{(C)}, \label{eq:2ptConditionC}\\
\delta_{V1}(3\delta_{3} + 6\delta_{V1} + 7) - \delta_{V2} =0, \hspace{13mm} \text{(A)}, \label{eq:2ptConditionA}
\end{eqnarray}
where we have defined $\delta_{V1(2)} \equiv \delta_{Z1(2)} = \pm \delta_{W1(2)}$ and the first (second) equation provides the condition for the custodial (anti-custodial) limit. Prima facie, this might appear like an added complication, as the Higgs mass appear to be quartically sensitive to the UV scale as opposed to the expected quadratic dependence at 1-loop; nevertheless, we show below that this would make the situation simpler.

The equivalence theorem provides us an insight into how to deal with this situation. Note that each $hGG$ vertex is associated with two derivatives. Thus, each Higgs insertion in a $G_{0,\pm}$ loop will increase the leading power of the cutoff scale $\Lambda$ by two. On the other hand, each new Higgs insertion will create an additional $G_{0,\pm}$ propagator $\sim 1/p^{2}$, which lowers the leading power by two, thereby canceling out the two additional powers that originate from the vertex. Therefore, not only will the 2-point function be quartically divergent, but also will all $N$-point functions. Thus, each $N$-point function will be associated with a separate condition to cancel out its quartic divergence.
\newline
\begin{figure}[H] 
  \centering
    \includegraphics[width=0.48\textwidth]{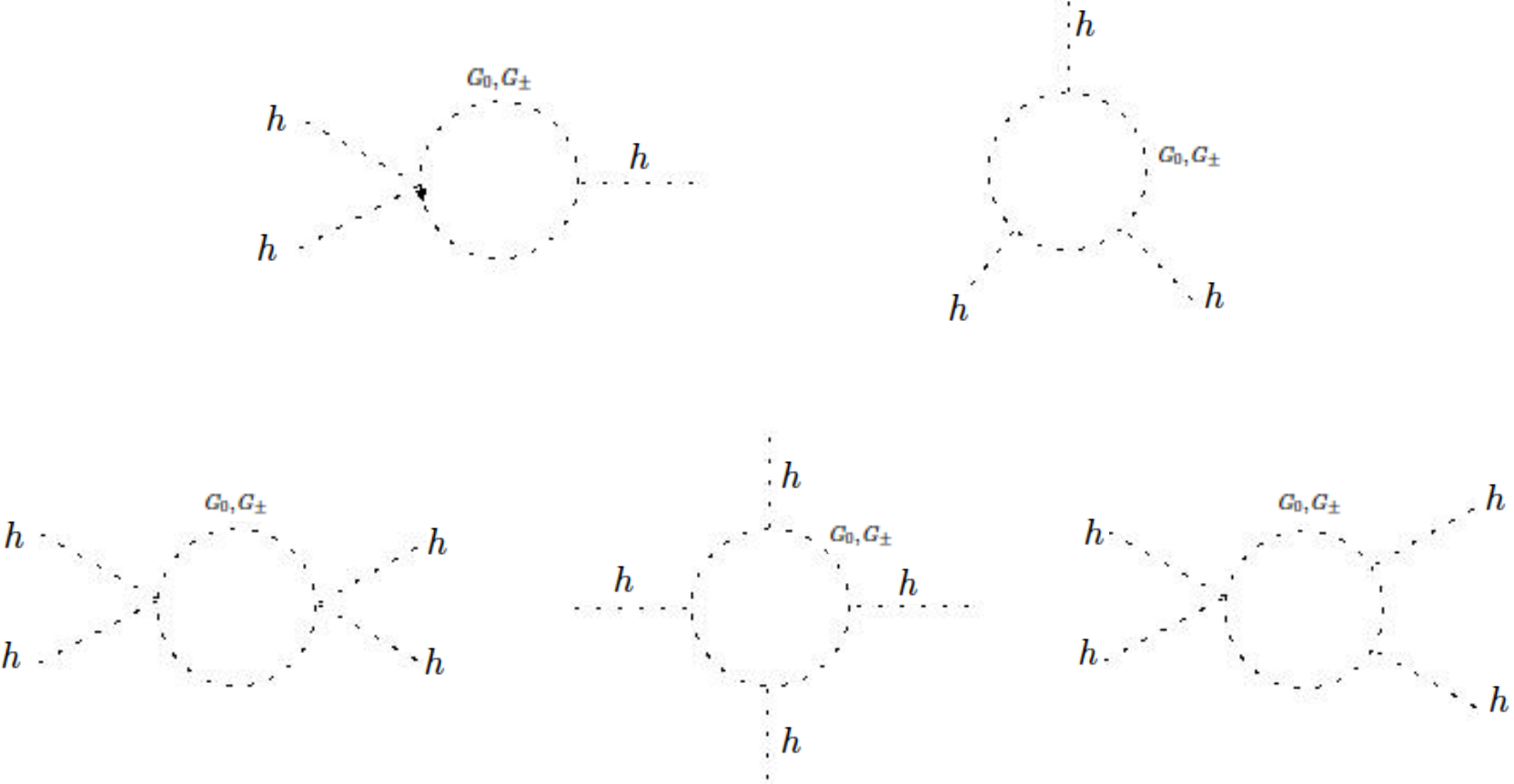}
      \caption{(Top): Feynman diagrams of the Higgs 3-point function with a longitudinal mode loop. (Bottom), Feynman diagrams for the Higgs 4-point function.}.
      \label{fig:Feynman3}
\end{figure}
The top row of Fig. \ref{fig:Feynman3} shows the 1-loop contributions to the Higgs 3-point function. A simple calculation yields the following conditions for canceling out the quartic divergences in the custodial and anti-custodial limits:
\begin{flalign}
& \delta_{V1}(9\delta_{V2}-36 \delta_{V1}+8\delta_{V1}^{2})=0, \hspace{17mm} \text{(C)},\label{eq:3ptConditionC} \\
& \delta_{V1}(\delta_{V2}-4\delta_{V1}) =0, \hspace{32mm} \text{(A)}.\label{eq:3ptConditionA}
\end{flalign}

Similarly, the bottom row of Fig. \ref{fig:Feynman3} shows the 1-loop contributions to the Higgs 4-point function and provides the following cancelation condition:
\begin{flalign}
& 16\delta_{V1}^{4} - 64 \delta_{V1}^{3} +16\delta_{V1}^{2}\delta_{V2} 
+48 \delta_{V1}^{2} \nonumber \\
& - 24\delta_{V1}\delta_{V2}+3\delta_{V2}^{2}=0, & \text{(C)},\label{eq:4ptConditionC} \\
& 16\delta_{V1}^{4}+48 \delta_{V1}^{2}-24\delta_{V1}\delta_{V2}+3\delta_{V2}^{2}=0, & \text{(A)}. \label{eq:4ptConditionA}
\end{flalign}

In order for the 2-, 3-, and 4-point functions to be free of quartic divergences, one has to solve Eqs. (\ref{eq:2ptConditionC}), (\ref{eq:3ptConditionC}), and (\ref{eq:4ptConditionC}) simultaneously for the custodial limit, or Eqs. (\ref{eq:2ptConditionA}), (\ref{eq:3ptConditionA}), and (\ref{eq:4ptConditionA}) for the anti-custodial limit. It can be observed that, in both limits, the only possible solution is to set $\delta_{V1} = \delta_{V2} =0$. 
\newline
\begin{figure}[!h] 
  \centering
    \includegraphics[width=0.25\textwidth]{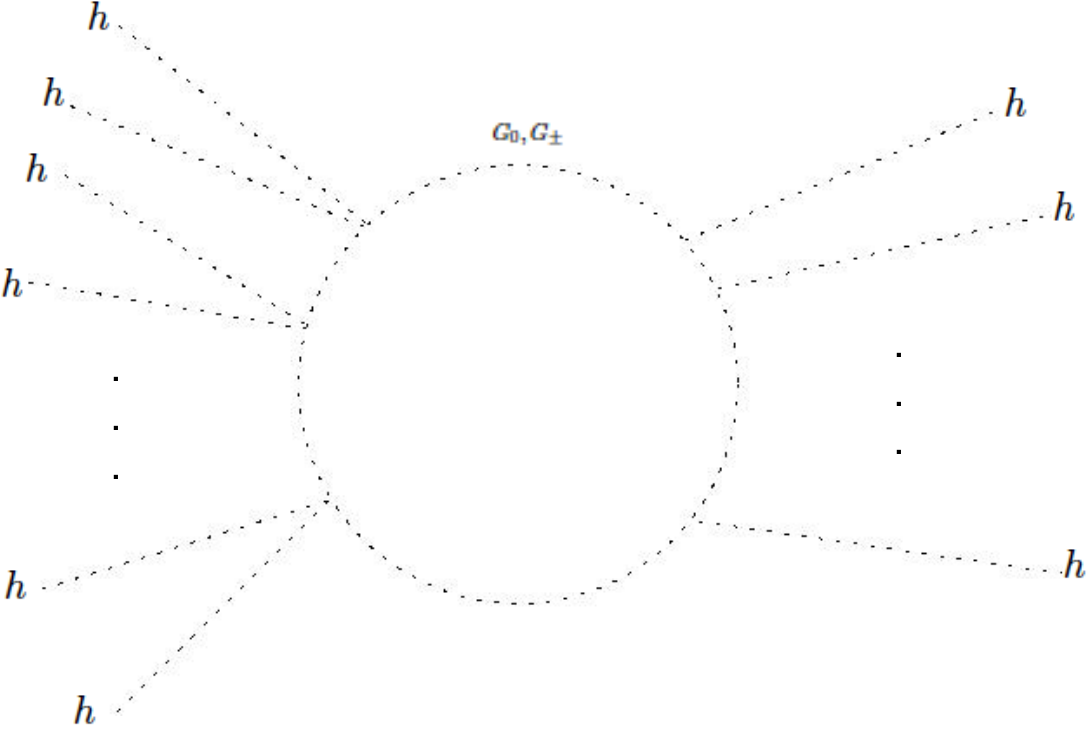}
      \caption{$G_{0}, G_{\pm}$ loop contribution to the Higgs $N$-point function with $n$ trilinear insertions and $m$ quartic insertions}
      \label{fig:Feynman4}
\end{figure}

We can sketch a general proof without making any assumptions about $\rho$ as follows: Consider a 1-loop $N$-point Higgs function with $G_{0}, G_{\pm}$ running in the loop, and with $n$ trilinear insertions and $m$ quartic ones, as shown in Fig \ref{fig:Feynman4}, such that $2m+n = N \in \mathbb{Z} > 2$, $n,m \in \mathbb{Z} \geq 0$. Then, schematically, the amplitudes are given by
\begin{equation}\label{eq:NG0amplitudes}
\mathcal{M}_{G_{0}}^{(m,n)} = \sum_{\substack{{m,n}\\2m+n=N\\N=3}}^{\infty} \frac{C_{m,n}^{0}}{16\pi^{2}v^{N}} \delta_{Z1}^{n} [\delta_{Z2} - 4\delta_{Z1}]^{m} \Lambda^{4},
\end{equation}
\begin{equation}\label{eq:NGpmamplitudes}
\mathcal{M}_{G_{\pm}}^{(m,n)} = \sum_{\substack{{m,n}\\2m+n=N\\N=3}}^{\infty} \frac{C_{m,n}^{\pm}}{16\pi^{2}v^{N}} \delta_{W1}^{n} [\delta_{W2} - 4\delta_{W1}]^{m} \Lambda^{4},
\end{equation}
where $C_{m,n}^{0}$ and $C_{m,n}^{\pm}$ are distinct rational numbers that could include possible symmetry factors. To avoid any quartic divergences in all $N$-point functions, one needs to impose the following infinite tower of linearly-independent conditions for the sum of amplitudes:
\begin{flalign}\label{eq:GeneralCondition}
\mathcal{M}_{G_{0}}^{(2)} + \mathcal{M}_{G_{\pm}}^{(2)} +  \mathcal{M}_{G_{0}}^{(m,n)}  +\mathcal{M}_{G_{\pm}}^{(m,n)} & =  0,
\end{flalign}
$\forall \hspace{1mm}  2m+n = N\hspace{1mm} > 2$, where $\mathcal{M}_{G_{0,\pm}}^{(2)}$ represent the contributions to the 2-point function given in Eqs. (\ref{eq:QuarticDiv-a})- (\ref{eq:QuarticDiv-f}). This infinite tower of conditions can only be satisfied if all the deviations vanish identically. Thus, we arrive at our first important result:
\begin{quote} 
\textit{Assuming that there are no degrees of freedom that couple to the Higgs below some UV scale $\Lambda$, then to avoid quartic divergences in the Higgs mass and all Higgs $N$-point functions with a longitudinal gauge boson loop, all the deviations in the couplings of the massive gauge bosons to the Higgs from the SM predictions must vanish.\footnote{A possible loophole in this result is if one assumes that corrections to $N$-point functions at one loop cancel against those from higher loops, however, such a scenario would require unnatural cancelations and fine-tuning, thus we ignore it here.}
}
\end{quote} 

However, we should note that, if $\delta_{V1,2}$ are small and if the scale of new physics is not too large, then the corrections to the Higgs mass and $N$-point functions could be maintained within acceptable limits; however, this would imply that the scale of new physics would be too low if the couplings of the Higgs to the $W$ and $Z$ were to have any meaningful deviations from the SM predictions. For example, from Eqs (\ref{eq:QuarticDiv-a} - \ref{eq:QuarticDiv-f}), we can estimate the contributions of the Goldstone bosons to the Higgs mass as
\begin{equation}\label{eq:GtoHiggsMass}
\delta m^{2} \sim \frac{9\delta_{V}}{16\pi^{2}v^{2}}\Lambda^{4},
\end{equation}
where we have assumed that $\delta_{V1} \sim \delta_{V2} = \delta_{V}$ and set $\delta_{3} =0$.\footnote{Non-vanishing values of $\delta_{3}$ don't affect the scale by much.} Then, for a deviation of approximately $1\%$ from the SM, the scale of new physics should be as low as $\sim 1$ TeV to yield corrections to the Higgs mass at $O(100\%)$, whereas a deviation of $O(10^{-4})$ can only push the scale of new physics up to $\sim 3.5$ TeV.  Such a low scale of new physics is in opposition to the null results from the LHC. If the scale of new physics is assumed to be larger than the energy scale of the LHC at $13$ TeV, then we can estimate the level of deviations that provide corrections to the Higgs mass at the $O(100\%)$ to be of $O(10^{-7})$. This result appears to suggest strongly that $\delta_{V1,2}$ should vanish for all practical purposes. This result is intuitive, as any (apparent) gauge violation in the EFT should be small if the scale of new physics is not too low.

Setting $\delta_{V1,2} = 0$, we can immediately calculate the remaining 1-loop corrections to the Higgs mass. We present the full results in Appendix \ref{app:MassCorrections}. Using these results, we readily extract the model-independent BSM Veltman condition
\begin{flalign}\label{eq:BSMVeltman}
& \big[3\delta_{3}^{2} + 6\delta_{3}-\delta_{4}+4 \big]m_{h}^{2} +2\big[3\delta_{3}+2 \big](2m_{W}^{2}+m_{Z}^{2}) \nonumber \\
& + 8m_{t}^{2} \big[\delta_{t1}^{2} + c_{t2}- \delta_{t1}-3\delta_{3}\delta_{t1} - 3\delta_{3} -2 \big]= 0.
\end{flalign}

Note that, when $\delta_{3}, \delta_{4}, \delta_{t1}, c_{t2} \rightarrow 0$, we retrieve the original Veltman condition in Eq. (\ref{eq:SM_Veltman}), as we should. Moreover, that this condition is completely model-independent, as it does not depend on any assumed UV completion and only depends on the deviations from the SM predictions. Any new physics at a UV scale $\Lambda$ should be matched to the EFT parameters in Eq. (\ref{eq:BSMVeltman}) at leading order once the heavy degrees of freedom have been integrated out. 

Satisfying the BSM Veltman condition would guarantee the cancelation of the quadratic divergences at 1-loop; however, there is no guarantee that higher-loop corrections would be canceled out and they probably may not. Nevertheless, we show in Sec. \ref{Sec:fine-tuning} that higher-order loops are not a concer, as they only provice corrections of the order to the Higgs mass, and thus, do not require fine-tuning.

\subsection{A Note on Dimensional Regularization}\label{Sec:dimreg}
It has been argued in the literature that the quadratic divergences in the Higgs mass are merely an artifact of using a cutoff regularization scheme, which breaks gauge invariance, and that using dimensional regularization, which \textit{is} gauge invariant, would remove any quadratic divergences, thus curing the hierarchy problem once the right counterterms have been subtracted; see for instance \cite{Manohar:2018aog}. However, as shown in the introduction, if there are heavy degrees of freedom that do couple to the Higgs, then after integrating these degrees of freedom out, the Higgs mass is quadratically sensitive to the UV mass scale of these degrees of freedom. Thus, fine-tuning cannot be explained away using dimensional regularization unless these heavy degrees of freedom are absent. More concretely, dimensional regularization will be the correct scheme to use only if there are no heavy degrees of freedom that couple to the Higgs, which is a dubious claim, to say the least.

The use of dimensional regularization to probe the scale of new physics is inadequate for two reasons. First, dimensional regularization mixes all types of divergences; it mixes UV divergences with IR ones, and mixes quadratic divergence with logarithmic ones, collecting all divergences in factors of $1/\epsilon$, which are canceled out by counterterms; and second, scaleless integrals of the form 
\begin{equation}\label{eq:scaleless}
\int_{0}^{\infty} \frac{d^{4}k}{(2\pi)^{4}}\frac{1}{k^{2n}},
\end{equation}
which arise from diagrams with Goldstone bosons running in the loop, vanish in dimensional regularization. However, if there is a UV scale $\Lambda$ where new physics comes into play, then one has to integrate up to that scale, and these integrals no longer vanish.

In other words, while dimensional regularization can yield finite results for the EFT, and thus, enable accurate quantitative calculations of known physics, it nonetheless cannot capture the scale of new physics, should such a scale exist.

An example of where dimensional regularization fails to predict new physics while a cutoff scheme succeeds is the GIM mechanism \cite{Glashow:1970gm}. In the GIM mechanism, quadratic divergences are canceled out by virtue of the unitarity of the CKM matrix; however, if one were to consider an effective theory where some of the heavy quarks are missing (or have been integrated out), then the GIM mechanism no longer cancels out the quadratic divergences, as the unitarity of the CKM matrix is lost, and any attempt at using dimensional regularization completely obscures new physics.

To illustrate this point further, let us imagine an EFT of only the $u$, $d$, and $s$ quarks and use it to study $s\bar{d} \rightarrow d\bar{s}$ processes, such as the mixing of $K_{L}^{0} - K_{S}^{0}$. Using a cutoff to calculate the 1-loop amplitude, one can find the $K_{L}^{0} - K_{S}^{0}$ mass splitting \cite{falkowski}
\begin{equation}\label{eq:Kmass}
\frac{\Delta m_{K^{0}}}{m_{K^{0}}} \sim \frac{\sin^{2}{\theta_{c}}\Lambda_{QCD}^{2}}{4\pi^{2}v^{4}}\Lambda^{2},
\end{equation} 
where $\theta_{c}$ is the Cabibbo angle. Given the measured values of $\Delta m_{K^{0}}$ and the kaon mass, one finds a scale of new physics of $\Lambda \lesssim 1$ GeV, which accurately predicts the existence and mass of the charm quark. On the other hand, using dimensional regularization, one would completely miss this prediction, as one would naively subtract the $1/\epsilon$ divergence, obtaining a finite result that does not point to the charm quark. The prediction and discovery of the charm completely exonerate the use of a cutoff scheme.

As apriori, we do not know whether there is any physics BSM that couples to the Higgs, and we are attempting to explore this potential new physics and the UV scale thereof, using a cutoff scheme is the appropriate prescription.
\subsection{Experimental Constraints and the Parameter Space}\label{Sec:constraints}
The BSM Veltman condition has the following parameters:
\begin{equation}\label{eq:BSMparameters}
\delta_{3}, \hspace{1mm} \delta_{4}, \hspace{1mm} \delta_{t1}, \hspace{1mm} c_{t2}.
\end{equation}

Within the $\kappa$ framework, $\delta_{i} = \kappa-1$, whereas $c_{t2}$ has no SM counterpart. The latest measurements from the LHC place the $95\%$ level bounds on $\delta_{3}$ and $\delta_{t1}$ at \cite{ATLAS:2018otd, Aad:2019mbh}
\begin{subequations}\label{eq:delta_bounds}
\begin{equation}
-6 \leq \delta_{3} \leq 11.1,
\end{equation}
\begin{equation}
-0.18 \leq \delta_{t1} \leq 0.24.
\end{equation}
\end{subequations}

On the other hand, $\delta_{4}$ remains unmeasured and thus unconstrained. In this study, we set the bounds on $\delta_{4}$ to be the same as those on $\delta_{3}$ as a conservative estimate. Bounds on $c_{t2}$ are more complicated, as they depend non-trivially on $\delta_{3}$ through the di-Higgs production \cite{Azatov:2015oxa}. Curve-fitting gives the following approximate bound at the $95\%$ confidence level
\begin{equation}\label{eq:ct2Bound}
0.04 \hspace{0.5mm} \delta_{3}^{2} + c_{t2}^{2} + 0.32 \hspace{0.5mm} \delta_{3} \hspace{0.5mm} c_{t2} - 0.3 \hspace{0.5mm} \delta_{3} - 1.52 \hspace{0.5mm} c_{t2} \leq 1.9.
\end{equation} 

Given the above constraints, a quick evaluation of the BSM Veltman condition reveals the following remarks:
\begin{itemize}
\item With the current bounds, satisfying the BSM Veltman condition, and thus canceling out the quadratic divergences, remains viable;
\item It is not possible to satisfy the condition if $\delta_{3}$ and $c_{t2}$ are both set to zero, i.e.; $\delta_{4}$ and $\delta_{t1}$ alone are insufficient to cancel iut the quadratic divergences,
\item It is possible for $\delta_{3}$ alone to satisfy the condition while all other deviations vanish. Similarly, $c_{t2}$ can also satisfy the condition with all other deviations vanishing, in which case $c_{t2}$ is constrained to be positive.
\end{itemize}

\begin{figure}[H]
\centering
\begin{tabular}{c c}
\subf{\includegraphics[width=0.22\textwidth]{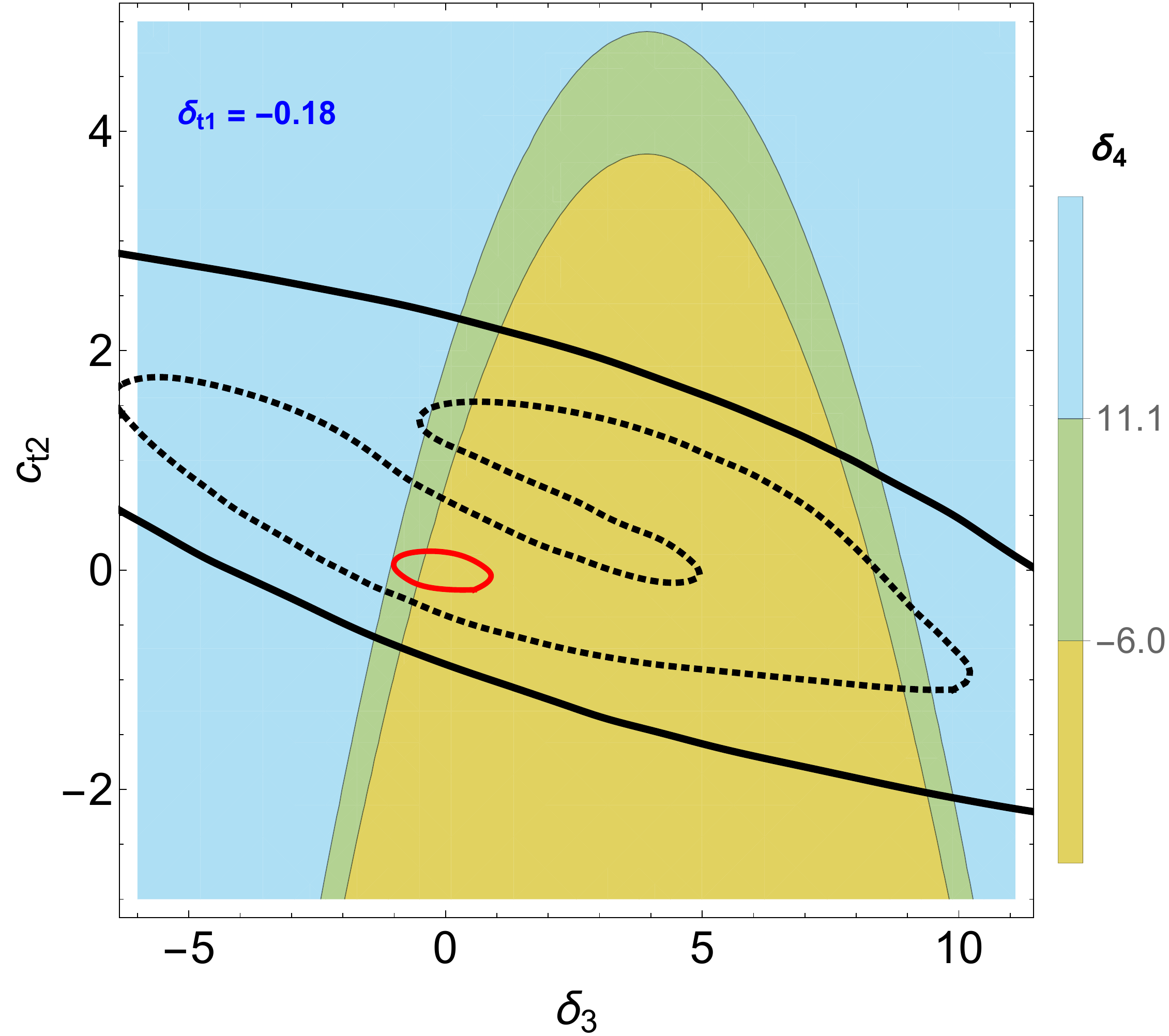}}{}
&
\subf{\includegraphics[width=0.22\textwidth]{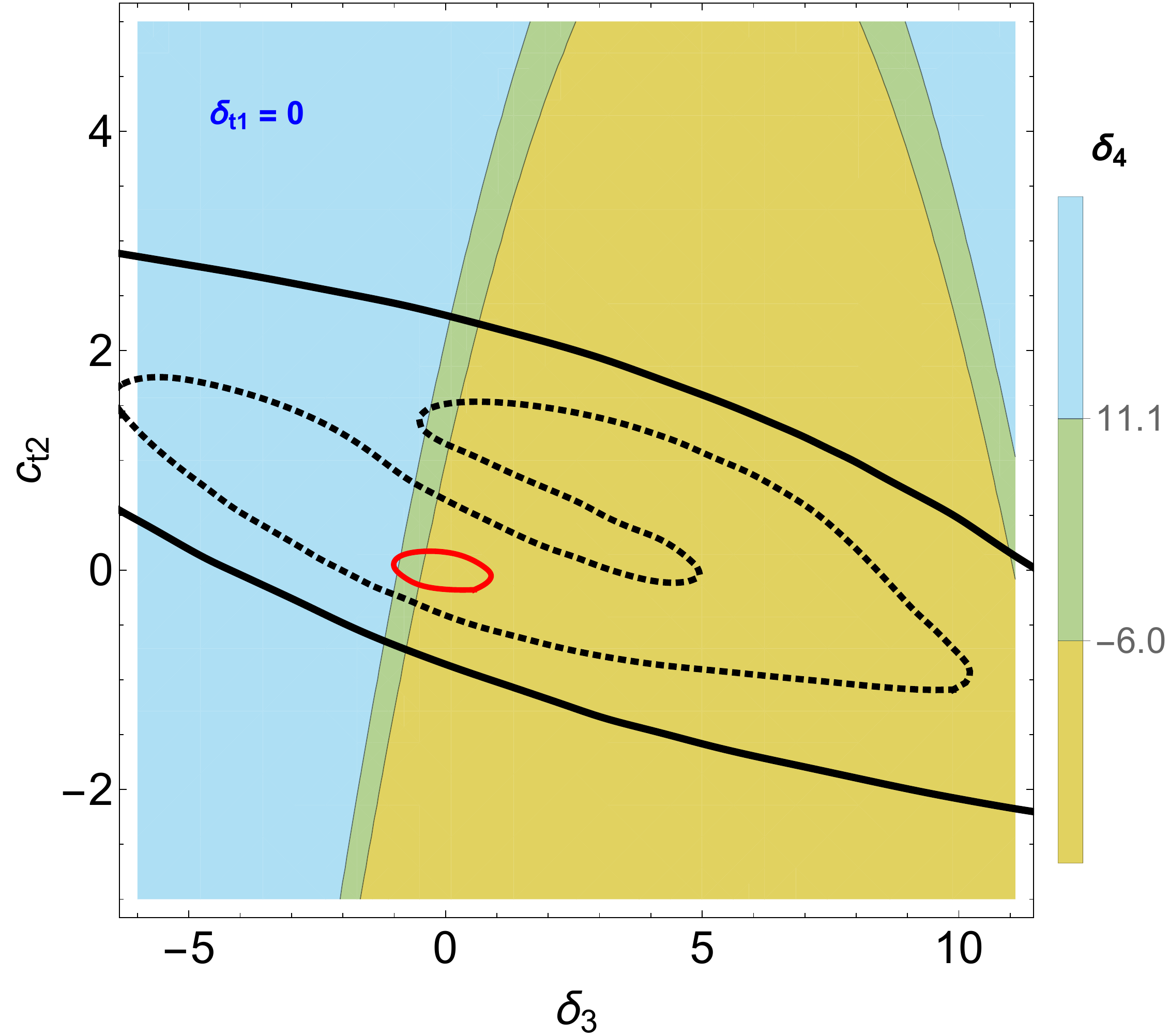}}{}
\\
\subf{\includegraphics[width=0.22\textwidth]{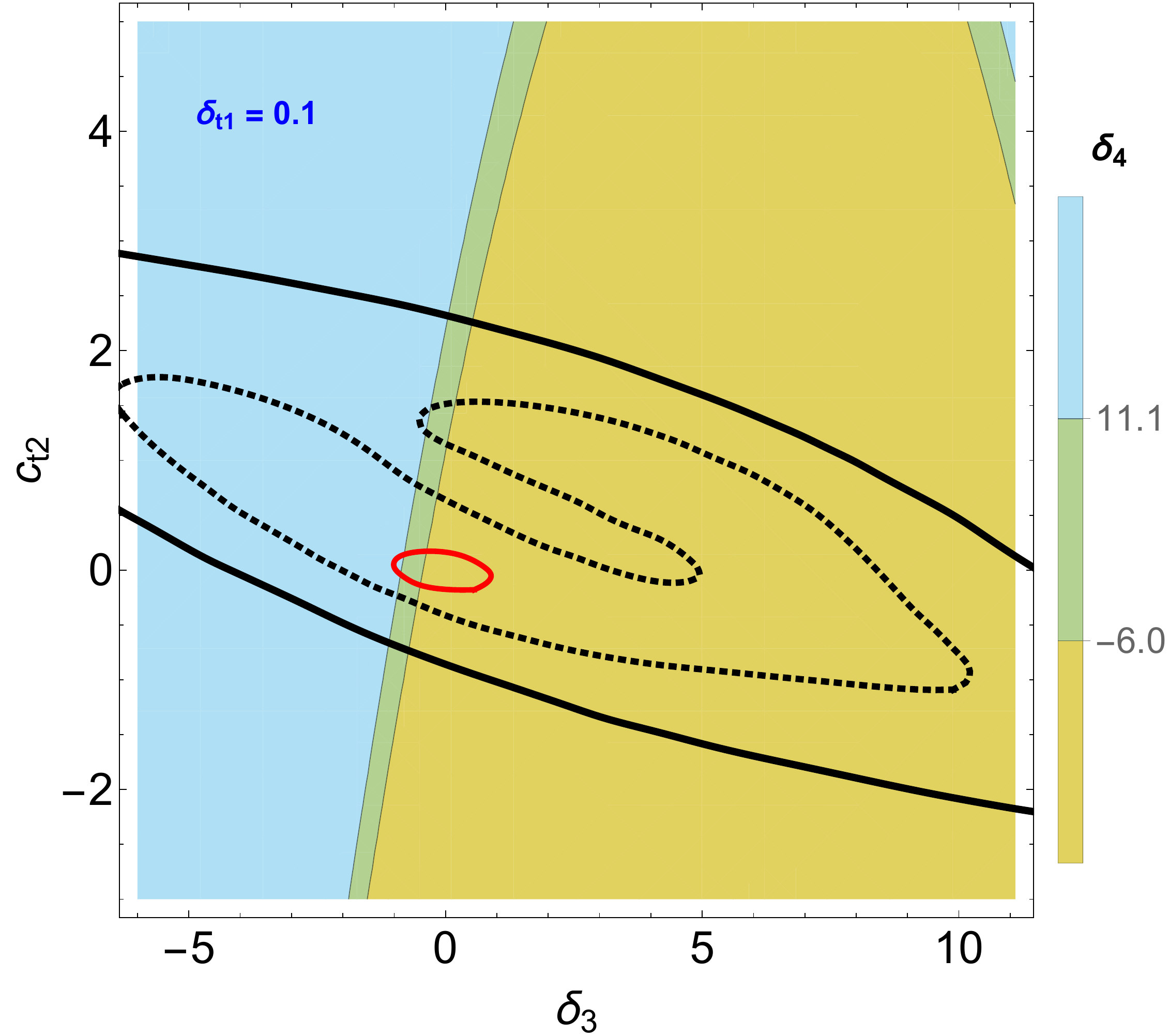}}{}
&
\subf{\includegraphics[width=0.22\textwidth]{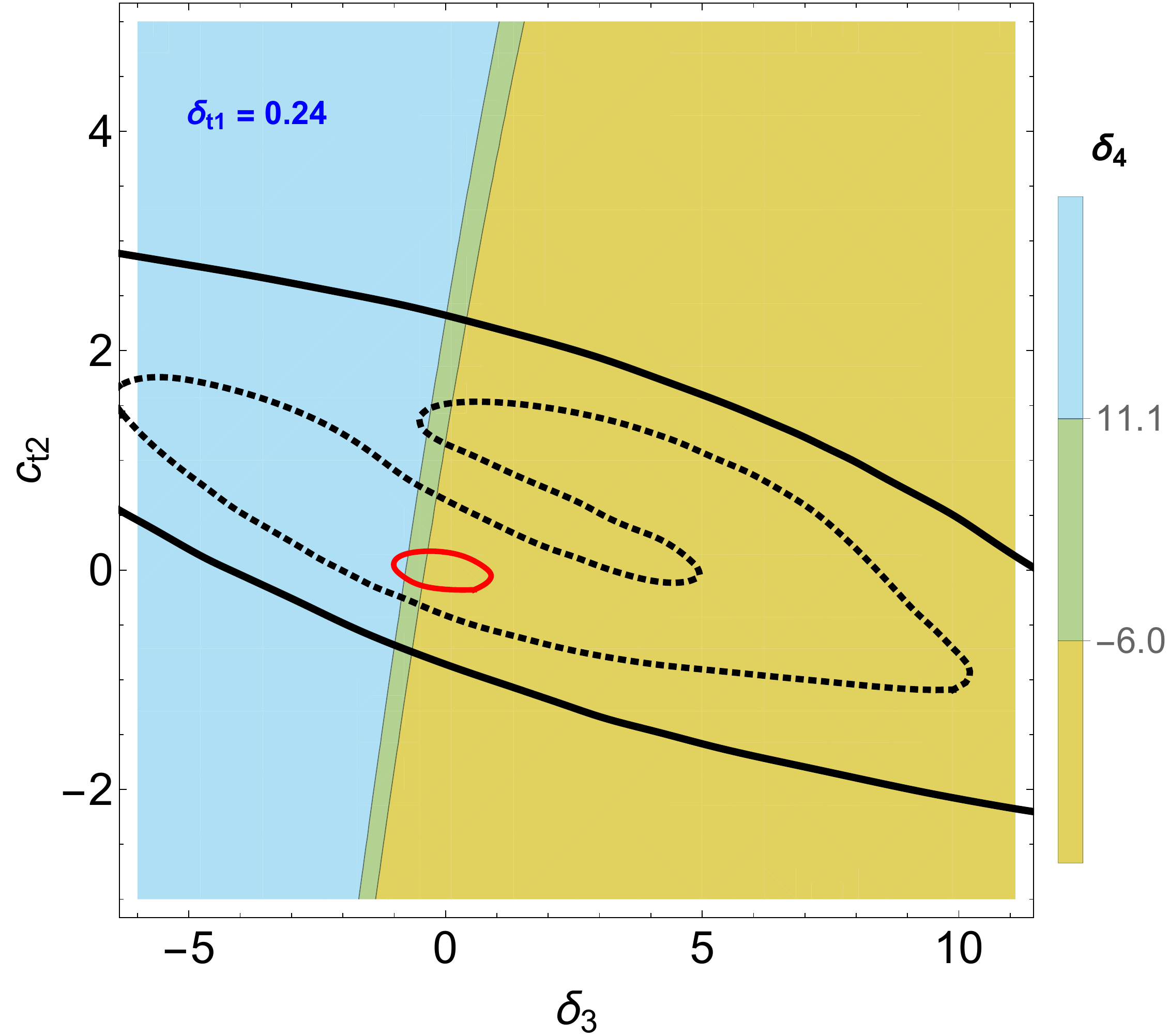}}{}
\\
\end{tabular}
\caption{Contour plots for $\delta_{3}$, $\delta_{4}$, and $c_{t2}$ for some benchmark values of $\delta_{t1}$. The plots also show the LHC limits (solid black line), along with the projections from the HL-LHC (dotted line) and the 100-TeV collider (solid red line). Only the green band that lies within the solid black contour is viable.}\label{fig:contours}
\end{figure}

In Fig. \ref{fig:contours}, we show the contour plots for $\delta_{3}$, $\delta_{4}$, and $c_{t2}$ for several benchmark values of $\delta_{t1}$. We superimpose the LHC constraints in addition to the projections of the High Luminosity LHC (HL-LHC) and the 100-TeV collider \cite{Azatov:2015oxa}. The viable region of the parameter space that can satisfy the BSM Veltman condition is the portion of the green band within the solid black contour. We can observe from the plots that the HL-LHC and 100-TeV colliders could probe much of the viable parameter space. Although the projection of the 100-TeV collider appears to suggest that the BSM Veltman condition could remain viable up to very high energies, and thus maintain the possibility of canceling out the quadratic divergences up to very large scales, we will show in Sec. \ref{Sec:Maximum_Energy} that the maximum energy scale at which the BSM Veltman condition remains viable is much lower. 
\newline

\section{Unitarity Violation and the Maximum Energy Scale}\label{Sec:Maximum_Energy}
The results of the previous section appear to suggest that the BSM Veltman condition could remain viable above $\sim 100$ TeV. While naively this might be the case if one were to inspect Eq. (\ref{eq:BSMVeltman}), we should note that deviations in the Higgs coupling will give rise to unitarity violating processes at the tree-level at a lower energy scale.

As pointed out in \cite{Chang:2019vez,Abu-Ajamieh:2020yqi}, the SM is the unique UV-complete theory with the observed particle content that can be extrapolated to arbitrarily high energy scales. This indicates that any deviation from the SM predictions would lead to UV incompleteness that manifests itself as energy-growing tree-level amplitudes that eventually violate unitarity at some high energy scale, signaling the onset of new physics BSM. This argument is along the same lines as the one presented by Lee, Quigg, and Thacker \cite{Lee:1977yc, Lee:1977eg}, which demonstrated the necessity of the existence of the Higgs boson itself with a mass below $\sim 1$ TeV. Following the argument in \cite{Chang:2019vez,Abu-Ajamieh:2020yqi}, one can set the unitarity condition as
\begin{equation}\label{eq:unitarity}
|\mathcal{\widehat{M}}| \leq 1,
\end{equation}
where $\mathcal{\widehat{M}}$ is the matrix element averaged over the initial and final phase spaces
\begin{equation}\label{eq:matrixElement}
\mathcal{\widehat{M}}_{fi}(P) = C_{f}^{*}C_{i} \int d\Phi_{f}(P) d\Phi_{i}(P) \mathcal{M}_{fi},
\end{equation}
and $C_{i,f}$ represent normalization constants. As $\mathcal{\widehat{M}} \sim \delta_{i}$, larger deviations indicate unitarity violation at lower energy scales, and as satisfying the BSM Veltman condition requires non-vanishing deviations, these deviations will lead to unitarity violation that points to a scale of new physics that becomes higher as the deviations become smaller. At a certain level, the size of the deviations will no longer be sufficient to satisfy the BSM Veltman condition. This indicates that there exists a maximum UV scale of new physics beyond which the quadratic divergences cannot be canceled out without significant fine-tuning. We seek to determine this energy scale.

Given the results in \cite{Chang:2019vez,Abu-Ajamieh:2020yqi}, the dominant unitarity violating processes in the Higgs sector are given in Table \ref{tab1}. We set the unitarity violating scale $E_{\text{max}}$ to be our UV cutoff $\Lambda$, and we use these processes to determine the maximum energy scale at which the BSM Veltman condition in Eq. (\ref{eq:BSMVeltman}) can still be satisfied, subject to the experimental constraints given in Eqs. (\ref{eq:delta_bounds}), and (\ref{eq:ct2Bound}). We find
\begin{equation}\label{eq:Emax}
E_{\text{max}}  = \Lambda \lesssim 19 \hspace{1mm} \text{TeV},\footnote{Notice that if we estimate the theoretical uncertainties in the unitarity violating amplitudes to be within a factor of $2$, as suggested in \cite{Chang:2019vez,Abu-Ajamieh:2020yqi}, then we can bound the scale of new physics to be within $ 13.5 \hspace{1mm} \text{TeV} \lesssim \Lambda \lesssim 27 \hspace{1mm} \text{TeV}$.} 
\end{equation}
at which $\delta_{3} \simeq -0.5$, $\delta_{4} \simeq -3$, $\delta_{t1} \simeq 0.03$, and $c_{t2} \simeq 0.09$, and unitarity violation originates from $G^{3} \rightarrow G^{3}$, $t_{R}\bar{b}_{R} \rightarrow G_{+}G_{-}G_{+}$, and $t_{R}\bar{b}_{R}G_{-} \rightarrow G_{+}G_{-}G_{+}G_{-}$. This indicates that, to avoid significant fine-tuning, any new physics that leads to the cancelation of the quadratic divergences in the Higgs mass should lie below $\sim 19$ TeV; otherwise, any UV contribution originating from integrating out the heavy degrees of freedom will not be sufficient to satisfy the BSM Veltman condition, and thus obtain a natural Higgs mass. To make this more concrete, we conjecture that
\begin{quote} 
\textit{if the deviations in the Higgs couplings $\delta_{3}$, $\delta_{4}$, $\delta_{t1}$, and $c_{t2}$ are too small to satisfy the BSM Veltman condition, or equivalently, if no new physics that couples to the Higgs boson is observed up to a scale of $\sim 19$ TeV, then either the Higgs sector is fine-tuned, or there is no UV sector that couples to the Higgs, possibly up to the Planck scale.\footnote{A possible loophole in this argument might be if one assumes that the Higgs couples to a UV sector whose various contributions cancel among themselves due to some UV symmetry, while the low energy contributions are only canceled by counterterms. However, such a scenario is somewhat unnatural.}
}
\end{quote}

This scale represents the Higgs little hierarchy. We discuss this point in more detail in the following section.
\begin{centering}
\begin{table}[!h]
\begin{tabular}{|c|c|} 
\hline
\bf{Process} & \bf{$E_{\text{max}}$ (TeV)} \\ [6pt]
\hline
$G^{3} \rightarrow G^{3}$ & $\frac{13.4}{\sqrt{|\delta_{3}|}}$ \\ [6pt]

 $h^{2}G_{0} \rightarrow hG_{0}$ & $\frac{66.7}{|\delta_{3} - \frac{1}{3}\delta_{4}|}$\\[6pt]

$hG_{0}^{2} \rightarrow hG_{0}^{2}$ & $\frac{9.1}{\sqrt{|\delta_{3}-\frac{1}{5}\delta_{4}|}}$  \\ [6pt]

$hG_{0}^{3} \rightarrow G_{0}^{3}$& $\frac{6.8}{|\delta_{3}-\frac{1}{6}\delta_{4}|^{1/3}}$\\[6pt]

$G_{0}^{4} \rightarrow G_{0}^{4}$ & $\frac{6.1}{|\delta_{3}-\frac{1}{6}\delta_{4}|^{1/4}}$ \\[6pt]

$t_{R}\bar{b}_{R} \rightarrow G_{+}G_{-}G_{+}$ & $\frac{3.33}{\sqrt{|\delta_{t1}|}}$ \\[6pt]

$\bar{t}_{R}t_{R} \rightarrow h^{2}$ & $\frac{7.18}{|c_{t2}|}$  \\[6pt]

$\bar{t}_{R}t_{R} \rightarrow G_{+}G_{-}h$& $\frac{4.7}{\sqrt{c_{t2}-2\delta_{t1}}}$\\[6pt]

$t_{R}\bar{b}_{R} \rightarrow G_{+}h^{2}$& $\frac{4.7}{\sqrt{c_{t2}-2\delta_{t1}}}$\\[6pt]

$t_{R}\bar{b}_{R}G_{-} \rightarrow h G_{+}G_{-}$ & $\frac{3.9}{|c_{t2}-3\delta_{t1}|^{1/3}}$\\[6pt]

$t_{R}\bar{b}_{R}W_{L}^{-} \rightarrow G_{+}G_{-}G_{+}G_{-}$ & $\frac{4.2}{|c_{t2}-3\delta_{t1}|^{1/4}}$\\[6pt]
\hline
\end{tabular}
\begin{minipage}{0.35\textwidth}
\caption{\small Dominant unitarity violating processes in the Higgs sector}
\label{tab1}
\end{minipage}
\end{table}
\end{centering}

\section{Mass Corrections, Fine-tuning and Naturalness}\label{Sec:fine-tuning}
In this section, we will discuss the fine-tuning associated with the BSM Veltman condition and estimate the mass corrections to the Higgs mass. The fine-tuning of the BSM Veltman condition depends on the scale of new physics $\Lambda$, which should be less than $\sim 19$ TeV. The scale of new physics depends in turn on the size of the deviations. To estimate the fine-tuning, we calculate the Barbieri-Guidice (BG) parameter \cite{Barbieri:1987fn} associated with the BSM Veltman condition
\begin{equation}\label{eq:BG}
\Delta = \text{Max} \Big\lbrace \Big| \frac{\delta_{i}}{m_{h}^{2}} \frac{\partial m_{h}^{2}(\delta_{3}, \delta_{4}, \delta_{t1}, c_{t2})}{\partial \delta_{i}} \Big| \Big \rbrace.
\end{equation}
\begin{figure}[!ht] 
  \centering
    \includegraphics[width=0.5\textwidth]{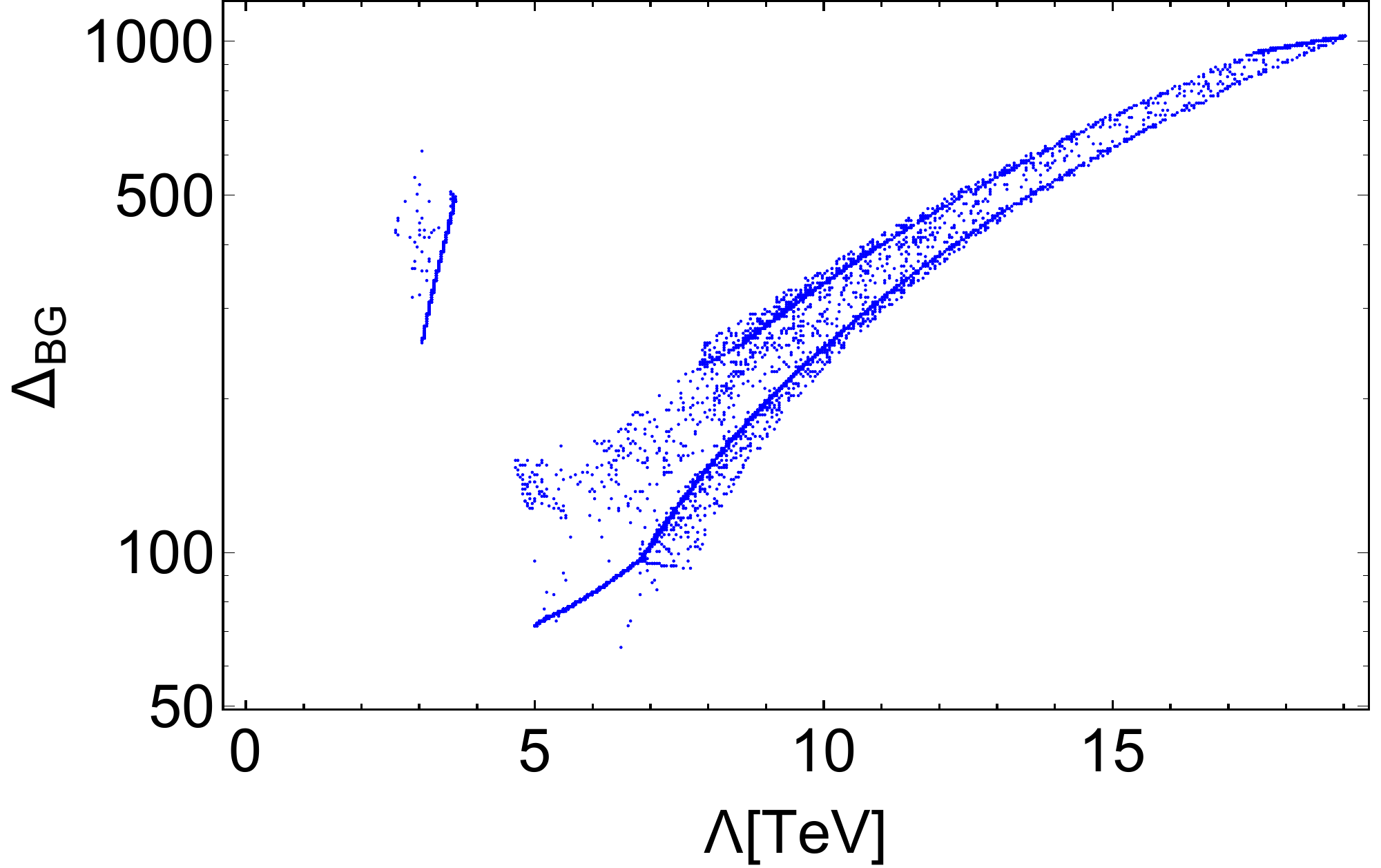}
      \caption{The BG parameter for the region of the parameter space that satisfies the BSM Veltman condition subject to the experimental constraints, plotted against the scale of new physics extracted from unitarity.}
      \label{fig:GB}
\end{figure}

We scan the viable parameter space that can satisfy the BSM Veltman condition in Eq. (\ref{eq:BSMVeltman}) subject to the constraints in Eqs. (\ref{eq:delta_bounds}) and (\ref{eq:ct2Bound}), and calculate the value of the BG parameter and the corresponding unitarity violating scale from the processes given in Table \ref{tab1}. We plot the BG parameter against the unitarity violating scale in Fig. \ref{fig:GB}. The plot shows that higher energy scales are associated with larger tuning, as expected. The values of the BG parameter range from as low as $\sim 60$, corresponding to a scale of new physics of $\sim 5$ TeV; to as high as $\sim 1000$, corresponding to a scale of $\sim 19$ TeV. This corresponds to a fine-tuning level of $O(1\%)$ - $O(0.1\%)$, respectively, which is indeed significant. This is to be expected as the BSM Veltman condition doesn't appear to possess any manifest symmetry that forces its satisfaction; however, if such a symmetry does exist in the UV sector, then this fine-tuning is eliminated.

More specifically, there are two possibilities for any UV sector that satisfies the BSM Veltman condition: 1) either the UV sector lacks any symmetry that forces the satisfaction of the BSM Veltman condition, in which case the quadratic divergences are canceled accidentally at the price of sub-percent fine-tuning, or 2) the UV sector is endowed with a symmetry that forces the satisfaction of the BSM Veltman condition and consequently the cancelation of the quadratic divergences naturally and without fine-tuning. 

Although the two scenarios of accidental cancelation and symmetry-based cancelation are equally viable solutions to the little hierarchy problem, the latter is more natural, and one tends to be more biased toward it. The BSM Veltman condition could provide us some insight as to what the UV completion should look like. We can observe by inspecting the condition that any proposed symmetry should relate the masses of the $W$, $Z$, top quark, and Higgs boson to the deviations in the Higgs couplings. Thus, any UV completion should provide corrections to the Higgs couplings that are related to these masses. One possibility would be to investigate if there is a viable supersymmetric realization of the BSM Veltman condition. We delay the issue of finding a symmetry-based UV completion to future work.
 
Assuming that such a symmetry-based UV theory exists, then the only fine-tuning that remains arises from the logarithmic and higher-order corrections to the Higgs mass. We can estimate the logarithmic corrections as
\begin{equation}\label{eq:logCorrections}
\delta m_{h}^{2} \sim \frac{m_{h}^{4}}{16\pi^{2}v^{2}}\log{\Big(\frac{\Lambda^{2}}{m_{h}^{2}} \Big)},
\end{equation} 
which for $\Lambda = 19$ TeV yields $\delta m_{h} \simeq 16$ GeV. On the other hand, we can use the Naive Dimensional Analysis (NDA) to estimate the NLO corrections
\begin{equation}\label{eq:NLOcorrections}
\delta m_{h}^{2} \sim \frac{m_{h}^{2}}{(16\pi^{2})^{2}v^{2}}\Lambda^{2},
\end{equation}
which yields $\delta m_{h} \simeq 61$ GeV for $\Lambda = 19$ TeV. Both corrections correspond to a fine-tuning level of $O(100\%)$, which is insignificant. Higher-order corrections will be further suppressed by additional loop factors of $1/16\pi^{2}$, and thus will be subleading and will not increase fine-tuning significantly. Thus, it will be sufficient to cancel out the quadratic divergences at 1-loop to eliminate fine-tuning and keep the Higgs mass natural.

\section{Conclusions}\label{Sec:conclusions}
In this study, we utilized a bottom-up EFT to derive a model-independent Veltman condition to cancel out the quadratic divergences in the Higgs mass. We first showed that using the equivalence theorem, all the deviations in the Higgs couplings to the $W$ and $Z$ from the SM predictions should vanish. We investigated the viable region of the parameter space where the BSM Veltman condition can be satisfied given the current constraints from the LHC, and we compared it with the projections of the HL-LHC and the 100-TeV collider.

We showed that, based on tree-level unitarity arguments, the highest energy scale at which the quadratic divergences can be canceled out should be $\sim 19$ TeV given the current level of uncertainty in the measurements of the Higgs couplings. Moreover, we argued that, to provide a natural solution to the little hierarchy problem, new physics should emerge at or below that scale. We conjectured that, if no new physics that couples to the Higgs is observed up to $\sim 19$ TeV, or equivalently, if the couplings of the Higgs to other SM particles are observed to be consistent with the SM predictions, then the Higgs either does not couple to any heavy degrees of freedom or is fine-tuned.

We discussed the aspects of naturalness and fine-tuning associated with the BSM Veltman condition and we observed that canceling out the quadratic divergences up to $\sim 19$ TeV is associated with sub-percent fine-tuning unless the UV completion has a symmetry that forces the satisfaction of the BSM Veltman condition, in which case any fine-tuning is eliminated. We showed in the latter case that logarithmic and higher loop corrections to the Higgs mass would not constitute any significant fine-tuning if the 1-loop contributions were to cancel out naturally. We postpone finding symmetry-based UV completions to future work. 

\section*{Acknowledgments}
I would like to thank Michele Frigerio and Gilbert Moultaka for the valuable discussions. I would also like to thank Markus Luty, Spencer Change, Felix Brummer, and Sacha Davidson. This work has been carried out thanks to the support of the OCEVU Labex (ANR-11-LABX-0060) and the A*MIDEX project (ANR-11-IDEX-0001-02) funded by the "Investissements d'Avenir" French government program managed by the ANR.
\appendix
\section{BSM Higgs Mass Corrections}\label{app:MassCorrections}
Here, we present the 1-loop contributions to the Higgs mass shown in Fig. \ref{fig:Feynman1} and \ref{fig:Feynman2}. In the Landau gauge $\xi = 0$, all ghost contributions vanish. The remaining contributions are (with $\delta_{Z1} = \delta_{Z2} = \delta_{W1} = \delta_{W2} =0$)
\begin{flalign}
\mathcal{M}_{h}^{(a)} & =  \frac{3(1+\delta_{4})m_{h}^{2}}{32\pi^{2} v^{2}}\Lambda^{2} + O\big( \log{\frac{\Lambda^{2}}{m_{h}^{2}}}\big), \\
\mathcal{M}_{G_{0}}^{(a)} & = \frac{(1+3\delta_{3})m_{h}^{2}}{32\pi^{2}v^{2}} \Lambda^{2}, \\
\mathcal{M}_{G_{\pm}}^{(a)} & = \frac{(1+3\delta_{3})m_{h}^{2}}{16\pi^{2}v^{2}} \Lambda^{2}, 
\end{flalign}
\begin{flalign}
\mathcal{M}_{h}^{(b)} & = -\frac{9(1+\delta_{3})^{2}m_{h}^{4}}{32\pi^{2}v^{2}} \log{\frac{\Lambda^{2}}{m_{h}^{2}}} + O(1), \\
\mathcal{M}_{G_{0}}^{(b)} & = -\frac{m_{h}^{4}}{32\pi^{2}v^{2}} \log{\frac{\Lambda^{2}}{m_{h}^{2}}}, \\
\mathcal{M}_{G_{\pm}}^{(b)} & = -\frac{m_{h}^{4}}{16\pi^{2}v^{2}} \log{\frac{\Lambda^{2}}{m_{h}^{2}}}, \\
\mathcal{M}_{t}^{(c)} & = -\frac{3(1+\delta_{t1})^{2}m_{t}^{2}}{4\pi^{2} v^{2}} \Lambda^{2} + O(\log{\frac{\Lambda^{2}}{m_{t}^{2}}}),\\
\mathcal{M}_{Z}^{(d)} & = -\frac{3m_{Z}^{4}}{8 \pi^{2} v^{2}}\log{\frac{\Lambda^{2}}{m_{Z}^{2}}},\\
\mathcal{M}_{W}^{(d)} & = -\frac{3m_{W}^{4}}{4 \pi^{2} v^{2}}\log{\frac{\Lambda^{2}}{m_{W}^{2}}},\\
\mathcal{M}_{Z}^{(e)} & = \frac{3m_{Z}^{2}}{16\pi^{2}v^{2}} \Lambda^{2} + O(\log{\frac{\Lambda^{2}}{m_{Z}^{2}}}),\\
\mathcal{M}_{W}^{(e)} & = \frac{3m_{W}^{2}}{8\pi^{2}v^{2}} \Lambda^{2} + O(\log{\frac{\Lambda^{2}}{m_{W}^{2}}}),\\
\mathcal{M}_{GZ}^{(f)} & =\mathcal{M}_{GW}^{(f)} = 0,\\
\mathcal{M}_{h}^{(h)} & = -\frac{9(1+\delta_{3})^{2}m_{h}^{2}}{32\pi^{2}v^{2}} \Lambda^{2} + O(\log{\frac{\Lambda^{2}}{m_{h}^{2}}}),\\
\mathcal{M}_{G_{0}}^{(h)} & = -\frac{3(1+\delta_{3})m_{h}^{2}}{32\pi^{2}v^{2}}\Lambda^{2},\\
\mathcal{M}_{G_{\pm}}^{(h)} & = -\frac{3(1+\delta_{3})m_{h}^{2}}{16\pi^{2}v^{2}}\Lambda^{2},\\
\mathcal{M}_{t}^{(i)} & = \frac{9(1+\delta_{3})(1+\delta_{t1})m_{t}^{2}}{4\pi^{2} v^{2}} \Lambda^{2} + O(\log{\frac{\Lambda^{2}}{m_{t}^{2}}}),\\
\mathcal{M}_{Z}^{(j)} & = -\frac{9(1+\delta_{3})m_{Z}^{2}}{16\pi^{2} v^{2}} \Lambda^{2} + O(\log{\frac{\Lambda^{2}}{m_{Z}^{2}}}),\\
\mathcal{M}_{W}^{(j)} & = -\frac{9(1+\delta_{3})m_{W}^{2}}{8\pi^{2} v^{2}} \Lambda^{2} + O(\log{\frac{\Lambda^{2}}{m_{W}^{2}}}),\\
\mathcal{M}_{t}^{(2)} & = -\frac{3c_{t2}m_{t}^{2}}{4\pi^{2}v^{2}} \Lambda^{2} + O(\log{\frac{\Lambda^{2}}{m_{t}^{2}}}).
\end{flalign}


\begin{thebibliography}{10}

\bibitem{Susskind:1978ms}
L.~Susskind,
Phys. Rev. D \textbf{20}, 2619-2625 (1979)

\bibitem{Susskind:1982mw}
L.~Susskind,
Phys. Rept. \textbf{104}, 181-193 (1984)

\bibitem{Veltman:1980mj}
M.~J.~G.~Veltman,
Acta Phys. Polon. B \textbf{12} (1981), 437
Print-80-0851 (MICHIGAN).

\bibitem{CapdequiPeyranere:1990gk}
M.~Capdequi Peyranere, J.~C.~Montero and G.~Moultaka,
Phys. Lett. B \textbf{260} (1991), 138-142

\bibitem{LopezCastro:1994mx}
G.~Lopez Castro and J.~Pestieau,
Mod. Phys. Lett. A \textbf{10}, 1155-1157 (1995)
\arXiv{9504350}{hep-ph}.

\bibitem{Bardeen:1995kv}
W.~A.~Bardeen,
FERMILAB-CONF-95-391-T.

\bibitem{Grange:2013yjn}
P.~Grange', J.~F.~Mathiot, B.~Mutet and E.~Werner,
Phys. Rev. D \textbf{88}, 125015 (2013)
\arXivold{1312.5278}{hep-ph}.

\bibitem{Aoki:2012xs}
H.~Aoki and S.~Iso,
Phys. Rev. D \textbf{86}, 013001 (2012)
\arXivold{1201.0857}{hep-ph}.

\bibitem{Vieira:2012ex}
A.~R.~Vieira, B.~Hiller, M.~C.~Nemes and M.~D.~R.~Sampaio,
Int. J. Theor. Phys. \textbf{52}, 3494-3503 (2013)
\arXivold{1207.4088}{hep-ph}.

\bibitem{Manohar:2018aog}
A.~V.~Manohar,
Les Houches Lect. Notes \textbf{108} (2020)
\arXivold{1804.05863 }{hep-ph}.

\bibitem{falkowski}
A.~Falkowski,
``Lectures on Effective Field Theories,''

\bibitem{Biswas:2020abl}
A.~Biswas, A.~Kundu and P.~Mondal,
Phys. Rev. D \textbf{102}, no.7, 075022 (2020)
\arXivold{2006.13513}{hep-ph}.

\bibitem{Chakraborty:2016izr}
I.~Chakraborty and A.~Kundu,
Pramana \textbf{87}, no.3, 38 (2016)

\bibitem{Chang:2019vez}
S.~Chang and M.~A.~Luty,
JHEP \textbf{03}, 140 (2020)
\arXivold{1902.05556}{hep-ph}.

\bibitem{Abu-Ajamieh:2020yqi}
F.~Abu-Ajamieh, S.~Chang, M.~Chen and M.~A.~Luty,
\arXivold{2009.11293}{hep-ph}.

\bibitem{Tanabashi:2018oca}
M.~Tanabashi \textit{et al.} [Particle Data Group],
Phys. Rev. D \textbf{98}, no.3, 030001 (2018)

\bibitem{Stolarski:2020qim}
D.~Stolarski and Y.~Wu,
Phys. Rev. D \textbf{102}, no.3, 033006 (2020)
\arXivold{2006.09374}{hep-ph}.

\bibitem{Glashow:1970gm}
S.~L.~Glashow, J.~Iliopoulos and L.~Maiani,
Phys. Rev. D \textbf{2}, 1285-1292 (1970)

\bibitem{ATLAS:2018otd}
 [ATLAS],
ATLAS-CONF-2018-043.

\bibitem{Aad:2019mbh}
G.~Aad \textit{et al.} [ATLAS],
Phys. Rev. D \textbf{101}, no.1, 012002 (2020)
\arXivold{1909.02845}{hep-ex}.

\bibitem{Azatov:2015oxa}
A.~Azatov, R.~Contino, G.~Panico and M.~Son,
Phys. Rev. D \textbf{92}, no.3, 035001 (2015)
\arXivold{1502.00539}{hep-ph}.

\bibitem{Lee:1977yc}
B.~W.~Lee, C.~Quigg and H.~B.~Thacker,
Phys. Rev. Lett. \textbf{38}, 883-885 (1977)

\bibitem{Lee:1977eg}
B.~W.~Lee, C.~Quigg and H.~B.~Thacker,
Phys. Rev. D \textbf{16}, 1519 (1977)


\bibitem{Barbieri:1987fn}
R.~Barbieri and G.~F.~Giudice,
Nucl. Phys. B \textbf{306}, 63-76 (1988)

\end{thebibliography}
\end{document}